\documentclass[aps,prb,preprint,amsmath,amssymb,showpacs]{revtex4}

\usepackage{graphicx}
\usepackage{dcolumn}
\usepackage{bm}
\usepackage{amsmath}
\usepackage{bm}
\usepackage{hyperref}

\newcommand{\ks} {{\bf k}}
\newcommand{\vs} {{\bf v}}

\newcommand{\ps} {{\bf p}}

\newcommand{\rs} {{\bf r}}
\newcommand{\es} {{\bf e}}
\newcommand{\ac} {{\bf A}}
\newcommand{\fc} {{\bf F}}
\newcommand{\gc} {{\bf G}}
\newcommand{\jc} {{\bf J}}
\newcommand{\eps} {{\varepsilon}}

\begin{document}

\title{Subcycle dynamics of electron-hole pairs and high-harmonic generation in bulk diamond subjected to intense femtosecond laser pulse}

\author{Tzveta Apostolova, Boyan Obreshkov}

\affiliation{Institute for Nuclear Research and Nuclear Energy,
Bulgarian Academy of Sciences, Tsarigradsko chausse\'{e} 72, Sofia
1784, Bulgaria}

\begin{abstract}

We present calculation of photoexcitation and  high-harmonic generation in bulk
diamond induced by  intense near-infrared laser pulse of photon energy 1.55 eV, time duration of 15 fs and peak field strength $F=$ 0.4 V/{\AA}.
Depending on the laser polarization direction, the pulsed irradiation   
creates electron-hole pairs and transient density fluctuations in a localized region of the crystal momentum space. 
As a consequence energetic inter- and intra-band harmonics are generated with alternating phase during each half-cycle of the driving pulse. 
The corresponding inter- and intra-band currents are in definite phase relation with the laser pulse, 
reflecting the build-up of coherent superposition of population between valence and conduction bands, 
and the accelerated motion of charge carriers in their respective bands.

\end{abstract}

\maketitle

\section{Introduction}

High-harmonic generation (HHG) is a non-linear process of laser frequency conversion to its multiples driven by the strong coupling of the laser field to electrons.
The laser-matter interaction results in coherent oscillation of the electrons and its periodicity is illustrated by emission of harmonics.
HHG is a way of producing intense attosecond pulses with wavelength extending from the XUV to
the soft X-ray region and the steps involved in high harmonic generation are shown to determine the time scale and energy of the generated pulses
\cite{xray,xuv,Corkum_2007,Krausz_Rev_Mod_Phys_2009}.
The application of single-cycle attosecond pulses in the VUV has enabled the tracing of electronic motion and electronic relaxation in atoms and molecules \cite{Christov_Opt_Comm_98}.
Femtosecond-laser-driven high harmonic generation sources have been used for coherent diffractive imaging (CDI) of biological samples and magnetic
domain patterns with nanoscale resolution \cite{Dinh_OC2017,Kfir_SA2017}. In situ characterization of the transient structural, optical,
and electronic properties of single nanoparticles has been achieved recently via CDI using XUV pulses from a HHG source \cite{Rupp_NatCom2017}.

The observation of nonperturbative high harmonic generation in a strongly driven bulk of wide-bandgap semiconductor
and dielectric materials  such as ZnO, GaSe and SiO$_2$ \cite{Ghimire_Nphys2011,Schubert_NatPhot2014, Ndabashimiye_Nat2015,Luu_Nat_2015} has enabled
the exploration of ultrafast electron dynamics in the condensed phase, laying the foundation of attosecond solid-state ultrafast spectroscopy \cite{Schultze_SCI2014}.
The generation of charged carriers via multiphoton or tunnel ionization at high intensities leading to dielectric breakdown has been carefully explored
to find regimes in which only HHG takes place avoiding material damage. Multiple plateaus  extending beyond the limit of the corresponding gas
phase harmonics\cite{Ghimire_Nphys2011} have been observed and linear scaling of the cutoff energy with the laser field has been found.
HHG sources from semiconductors and dielectrics have the potential of being brighter owing to the high atomic density of the target material
and the ability to withstand the driving laser fields necessary for inducing strong nonlinear and non perturbative effects underlying the production of high harmonic radiation.
Furthermore ultrafast nonlinear transient reversible electric currents below damage threshold
can be experimentally generated in dielectrics \cite{Schiffrin_Nat2013}, with  potential applications in modern electronics and petahertz-scale signal processing \cite{Krausz_NPhot2014,optica2016}.
Numerical simulations based on time-dependent density functional theory gave insight into atomic-scale properties of the induced ultrafast currents  \cite{Wachter_PRL2014}.
These results demonstrate that intense laser fields distort the electronic band structure and transform an insulator into
a metal on a sub-femtosecond
time scale.

Theoretical models have been proposed to analyze HHG in crystalline solids, using first principle approaches
\cite{Otobe_JCTN2009,Otobe_JAP2012,Otobe_PRB2016,Floss_2017,Tancogne-Dejean_PRL2017}, time-dependent-Schrodinger-equation (TDSE) \cite{Higuchi_PRL2014,Wu_PRA2015},
semiconductor Bloch equations \cite{Schubert_NatPhot2014,Luu_PRB2016} and semiclassical re-collision models  \cite{Vampa_cutoff}.
The theory has identified two main mechanisms of HHG in solids, based on the distinction between physical effects associated with intra- and interband motion
of photoexcited carriers, though in actual computation of HHG spectra these contributions are entangled \cite{gauge}.
The relative importance of these two mechanisms was investigated, cf. Ref.\cite{Vampa_cutoff}, 
over a wide range of wavelengths. For mid-IR lasers, generation of harmonic photons is dominated by interband mechanism, in contrast,
for far-infrared incident laser
wavelengths,  emission arising from intraband motion with nonlinear band velocities
(i.e. non-parabolic energy-momentum dispersion law) becomes dominant. Similarly to HHG in the gas phase \cite{Lewenstein_PRA_1994,Corkum_PRL_1993},
the interband emission in solids can be modelled semi-classically in terms of trajectories of electron-hole pairs in the laser electric field.
The outcome is an approximate law for the harmonic cutoff energy in solids showing that the maximum cutoff energy for HHG is limited by the
maximum band gap that can be reached by an electron-hole pair during one optical cycle of the laser field.
Recently HHG in model one-dimensional crystals  was studied numerically \cite{Ikemachi2017,Hansen2017}.  A
simplified momentum space analogue of the semi-classical three-step model for HHG in the gas phase was proposed in Ref.\cite{Ikemachi2017} by including tunnel ionization into the conduction band,
acceleration of electrons and valence band holes in their respective bands followed by their radiative recombination with emission of a single short wavelength photon . 
Importantly, such type of models incorporate the multi-band structure of the solid,
and successfully reproduce the experimentally observed multi-plateau structure of HHG in rare-gas solids (cf. also \cite{Wu_PRA2015,Hawkins}).
However reduced dimensionality models do not incorporate realistic three-dimensional static band-structures, moreover the crystal-momentum-dependence of 
couplings between valence and conduction bands is rarely discussed in the context of HHG.  Also numerical approaches are not well suited 
to expose laser-induced band-structure modification, that is fundamental in laser-solid interactions and HHG \cite{Faisal_pra1997,Gruzdev_prb2018}.

In the present work we focus on specific aspects of HHG associated with subcycle dynamics of photoexcited electron-hole pairs
in diamond bulk driven by linearly polarized intense 15 fs laser pulse with 800nm wavelength in detail.
The paper is organized as follows: Section II presents the theoretical method for modelling photo-excitation and photoionization
of diamond bulk. Section III is divided into three sub-sections: the first one includes description of the
static band structure of diamond and the crystal-momentum dependence of the transition dipole matrix element;
the second one presents some analytic results and analysis of HHG within a simplified two-band model, the third subsection
includes numerical results of a full multiband model that unreval the relative importance of interband and intraband dynamics of photoexcited
carriers in diamond.  Sec. IV includes our main conclusions.  Atomic units are used $e=\hbar=m_e=1$ throughout this paper.

\section{Pseudopotential method}

We describe the electron dynamics in a unit cell of crystalline diamond subjected to spatially uniform laser electric field  $\fc(t)$
with the time-dependent Schr\"{o}dinger equation in single-active electron approximation
\begin{equation}
i \partial_t |\psi_{v\ks} (t) \rangle= H(t)
|\psi_{v\ks}(t) \rangle, \label{tdse}
\end{equation}
where $v \ks$ labels the initially occupied valence band states of definite crystal momentum $\ks$ and
\begin{equation}
H(t) = \frac{1}{2}[\ps+\ac(t)]^2+V(\rs), \label{ham}
\end{equation}
is the time-dependent Hamiltonian in velocity gauge with $\fc(t)=-\partial_t \ac(t)$, here $\ac(t)$ is the laser vector potential,
$\ps=-i \nabla_{\rs}$ is the momentum operator and $V(\rs)$ is the periodic ion-lattice potential. In the local empirical pseudopotential method \cite{Cohen1966}, the scalar potential is
presented in a plane wave expansion in the basis of reciprocal lattice wave-vectors $\gc$
\begin{equation}
 V(\rs)=\sum_{\gc} V(G) \cos(\gc \cdot \boldsymbol \tau) e^{i \gc \cdot \rs}, \label{vps}
\end{equation}
here 2$\boldsymbol{\tau}=a_0(1/4,1/4,1/4)$ is a relative vector
connecting two carbon atoms in a crystal unit cell and $a_0=3.57$ {\AA} is the bulk lattice constant.
In Eq.(\ref{vps}), the pseudopotential formfactors $V(G)$ (in Rydberg)
are $V (G^2 = 3) = -0.625, V (G^2 = 8) = 0.051$, and $V (G^2 =
11) = 0.206$; here the wave number $G$ is measured in units of $2 \pi/a_0$.

To solve Eq.(\ref{tdse}) we use a basis set expansion over static Bloch states $|n \ks \rangle$
with associated eigenenergies $\eps_{n \ks}$, which are  solutions of the field-free Schr\"{o}dinger equation
$[\ps^2/2+V]|n \ks\rangle = \eps_{n \ks} | n \ks \rangle$, i.e. we write the expansion
\begin{equation}
|\psi_{v \ks}(t) \rangle = \sum_n a_{nv}(\ks,t) |n \ks \rangle \label{psik}.
\end{equation}
Substiuting Eq.(\ref{psik}) into Eq.(\ref{tdse}) gives a set of coupled equations for the transition amplitudes
\begin{equation}
\label{eom}
i \partial_t a_{nv}(\ks,t)=\eps_{n \ks} a_{n v}(\ks,t) + \sum_{n'} \ac(t) \cdot \ps_{nn'}(\ks) a_{n'v} (\ks,t),
\end{equation}
which are subject to initial conditions specified in the remote past $a_{nv}(\ks,t \rightarrow -\infty) =
\delta_{n v}$; In Eq.\ref{eom}, the state-independent field term $\ac^2(t)$ was absorbed into redefintion of the phases of basis states 
$|n \ks \rangle \rightarrow |n \ks \rangle \exp(-i \int^t dt' \ac^2(t'))$. The matrix representation of the momentum operator is
\begin{equation}
\ps_{nn'}(\ks)= \ks \delta_{nn'} +\sum_{\gc} \gc c^{\ast}_{n,\gc+\ks} c_{n',\gc+\ks}
\end{equation}
where $c_{n, \gc+\ks}$ are the Fourier coefficients in the expansion of the Bloch orbitals over plane waves.
The laser vector potential was assumed to be a temporally Gaussian pulse $\ac(t)=\es \exp(-ln(4) t^2/\tau^2) \sin \omega_L t$, where $\hbar \omega_L$=1.55 eV is the photon energy,
$\tau=$ 15 fs is the pulse duration and $\es$ is a unit vector in the direction of polarization.
The relevant output of the calculation are the time-dependent occupation numbers of Bloch states
\begin{equation}
f_{n \ks}(t) = \sum_v |a_{n v\ks}(t)|^2,
\end{equation}
the  transient distribution function  of conduction electrons
\begin{equation}
 f_e(\ks,t)= \sum_c f_{c \ks}(t),
\end{equation}
and the total number of electrons  at time $t$
\begin{equation}
n_e(t)= \int_{{\rm BZ}} \frac{d^3 \ks}{4 \pi^3} f_e(\ks,t).
\end{equation}
The instantaneous absorbed energy in a unit cell of the diamond lattice is
\begin{equation}
\Delta E (t)= \sum_v \int_{{\rm BZ}}
\frac{d^3\ks}{4\pi^3}\left\langle \psi_{v\ks}(t)  | H(t) | \psi_{v\ks} (t) \right\rangle -E_0,
\end{equation}
where $E_0$ is the ground state energy. After the end of the pulse ($t \rightarrow \infty$),
the  excitation energy relative to the ground state energy can be expressed by change of occupation of single-particle states
\begin{equation}
\Delta E (\infty) = \sum_n \int_{{\rm BZ}} \frac{d^3\ks}{4\pi^3} \eps_{n \ks} [ f_{n \ks} (\infty) - f^0_{n \ks} ]
\end{equation}
where $f^0_{v \ks}=1$ and $f^0_{c \ks}=0$.  In the electron-hole picture  with $f_{c \ks}=f_{e \ks}, f_{v \ks}=1-f_{h -\ks} ,
\eps_{h \ks}= -\eps_{v \ks}$ and  $\eps_{c \ks}= \eps_{e \ks}$, the absorbed energy $
\Delta E (\infty) =  \sum_{h, \ks} f_{h  \ks} \eps_{h \ks} + \sum_{e, \ks} f_{e \ks} \eps_{e \ks}$
is due to production of electron-hole pairs.

The optically-induced electron current density is given by the BZ integral
\begin{equation}
\jc(t)=\int_{{\rm BZ}} \frac{d^3 \ks}{4 \pi^3} \jc(\ks,t)
\end{equation}
where
\begin{equation}
\label{j}
\jc(\ks,t)=\sum_v \langle \psi_{v\ks}(t) | \vs(t) |
  \psi_{v\ks}(t) \rangle ,
\end{equation}
and  $\vs(t)=\ps+\ac(t)$ is the velocity operator. From the time-dependent current density,
the spectrum of high order harmonics can be calculated as a coherent sum
\begin{equation}
I(\omega)=\left|\int_{{\rm BZ}} \frac{d^3 \ks}{4 \pi^3}
J_{\ks,\omega} \right| \label{hhg_spec}
\end{equation}
over contributions from Bloch states with quasi-momentum $\ks$
\begin{equation}
J_{\ks,\omega}=\int dt e^{i \omega t} \es \cdot \jc(\ks,t).
\end{equation}

\section{Numerical Results and Discussion}

\subsection{Static band structure and matrix elements}

\begin{figure}
\begin{center}
\includegraphics[width=.5\textwidth]
{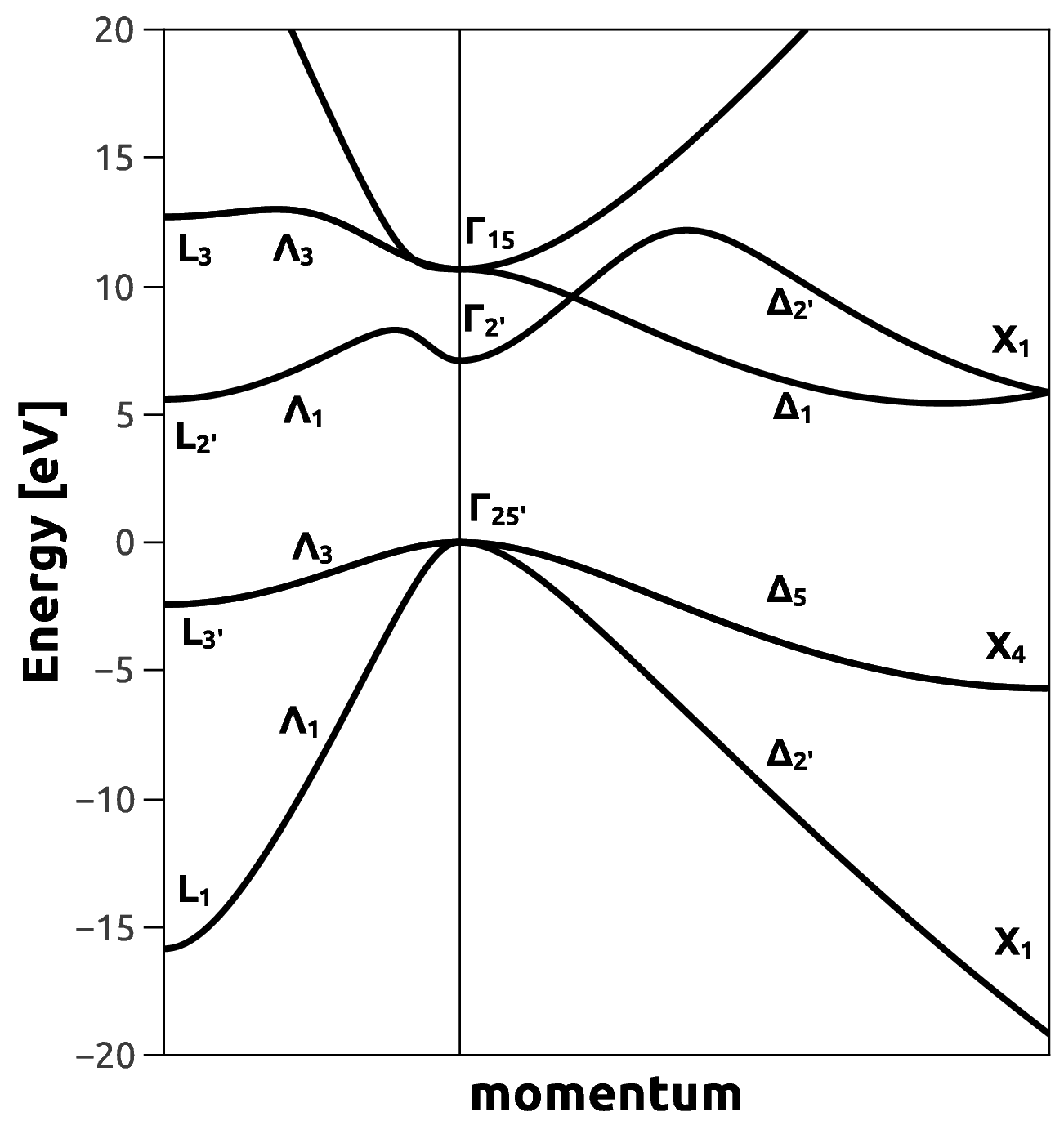} \caption{Electronic energy bands of diamond along the $\Delta$ and $\Lambda$ symmetry lines.} \label{fig01}
\end{center}
\end{figure}

The static energy-band structure of diamond along the $\Delta$ and $\Lambda$ lines is shown in Fig.\ref{fig01}.
The pseudopotential model reproduces quantitatively the principal energy gaps and the optical properties of diamond in the UV region \cite{Papadopoulos_prb1991}.
The location of the conduction band minimum at $\ks=(0.8,0,0)$ relative to the valence band maximum at $\Gamma$, specifies
the threshold for indirect tranistions 5.42 eV. The prominent direct bandgap tranistions are
$\Gamma_{25'} \rightarrow  \Gamma_{2'}$ (7.1 eV), $\Gamma_{25'} \rightarrow \Gamma_{15}$ (10 eV),
$L_{3'} \rightarrow L_{2'}$ (8 eV) , $L_{3'} \rightarrow L_3$ (13.1 eV), $X_4 \rightarrow X_1$ (11.5 eV).
The threshold for direct transitions  is assigned to the $\Gamma_{25'} \rightarrow \Gamma_{2'}$
energy gap and the main UV absorption peak can be assigned to the $X_1 \rightarrow X_4$ interband transition.
An important difference with regard to other band structure calculations is the energy level ordering at $\Gamma$ and L:
the pseudopotential parameters used in our calculation predict $\Gamma_{2'}$ lower in energy than $\Gamma_{15}$,
as a consequence $L_{2'}$ is also lower in energy than the $L_1$ and $L_3$ conduction bands (cf. also Ref.\cite{diamond_prb1970}).

\begin{figure}
\begin{center}
\includegraphics[width=.5\textwidth]
{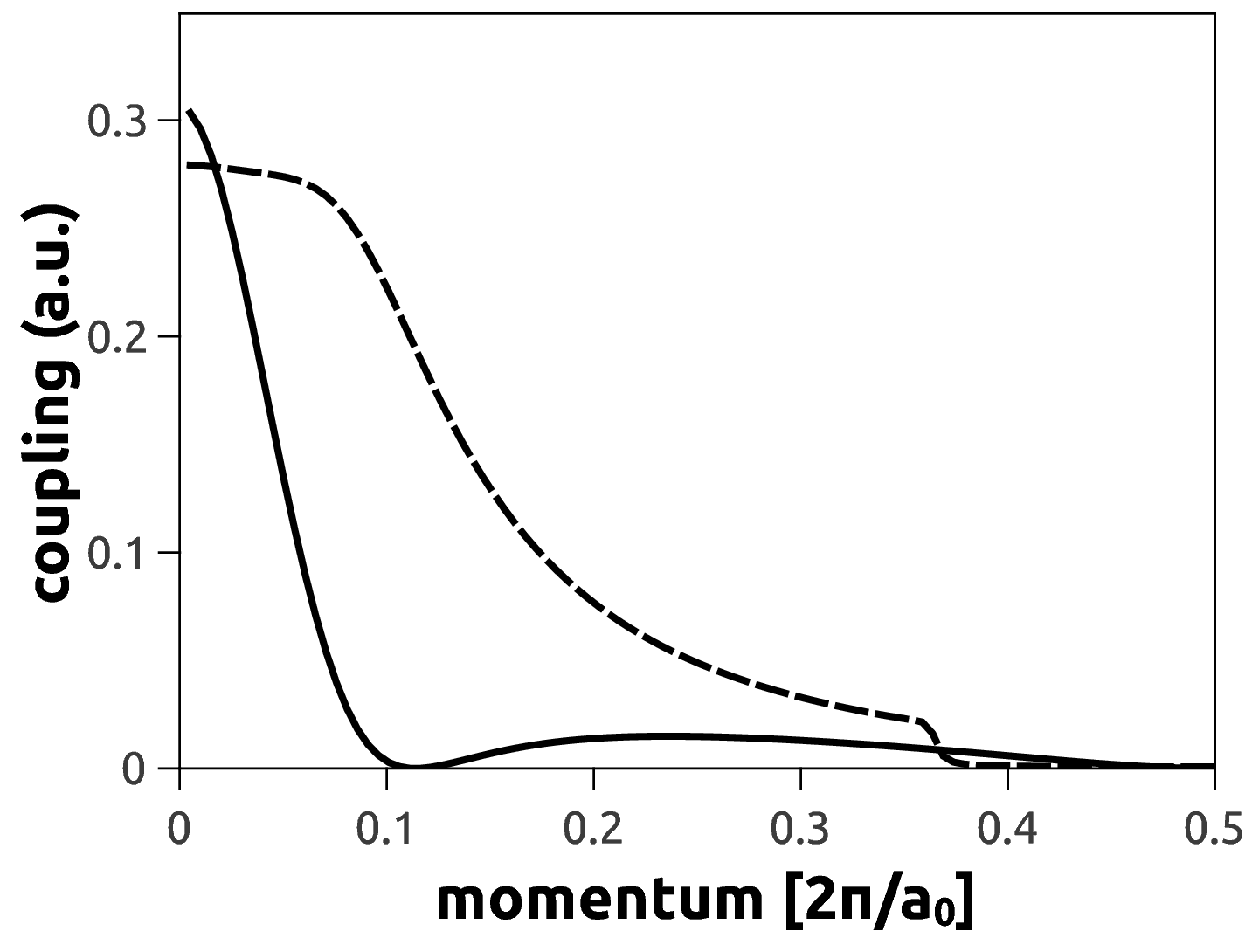} \caption{Squared matrix elements of the component of the momentum onto the laser polarization direction
between the light hole band and the lowest two conduction bands as a function of the crystal momentum $\ks=k(1,1,1)$:
$|\langle \Lambda_{1,lh}| \es \cdot \ps|  \Lambda_{1,c} \rangle|^2$ (solid line) and
$|\langle \Lambda_{1,lh}| \es \cdot \ps|  \Lambda_{3,c} \rangle|^2$ (dashed line). The laser polarization vector $\es$ is
pointing along the [111] direction. } \label{fig02}
\end{center}
\end{figure}

In Fig.\ref{fig02} we show the squared matrix elements $|\langle \Lambda_{1,lh} | \es \cdot \ps | \Lambda_{1,c} \rangle|^2$ and
$|\langle \Lambda_{1,lh} | \es \cdot \ps | \Lambda_{3,c} \rangle|^2$ between the
light hole and the lowest two conduction bands for laser linearly polarized along the [111] direction (cf. also Fig. \ref{fig01}).
For this specific laser polarization, the couplings among the heavy hole and these conduction bands vanish
$\langle \Lambda_{3,hh} | \es \cdot \ps | \Lambda_{1,c} \rangle=\langle \Lambda_{3,hh} | \es \cdot \ps | \Lambda_{3,c} \rangle=0$.
Noticeably the interactions with the light hole exhibit sensitive momentum dependence and are well localized  near the BZ center,
i.e. the $\Gamma_{25'} \rightarrow \Gamma_{2'}$ and $\Gamma_{25'} \rightarrow \Gamma_{15'}$ interband transitions should give dominant
contribution for the photoexcitation and HHG. It is worth noting that some authors assume constant value for the interband momentum matrix element \cite{hhg_semi},
which may be a good approximation for semiconductors \cite{Cardona_kp}, however we find that the detailed momentum-dependence of the interband couplings
should be accounted for HHG in widebandgap solids such as diamond.

\subsection{Time-dependent interactions}

\subsubsection{Two-band model} 

To examine the relevance of interband couplings and field-induced distortion of the diamond band structure,
we develop a simplified two band model describing the interactions between the light hole and
the lowest conduction band $M_k=\langle \Lambda_{1,c}  | \es \cdot \ps | \Lambda_{1,lh} \rangle$  for laser
irradiation linearly polarized along the [111] direction. For this specific direction in the BZ, $\ks=k(1,1,1)$,
the model Hamiltonian is
\begin{equation}
 H(k,t)= \frac{\Delta_k(t)}{2} (1+\tau_3) + V_k(t) \tau_1  
\end{equation}
where $(\tau_1,\tau_2,\tau_3)$ are the three Pauli matrices, 
$\Delta_k(t)=\Delta_k+v_k A(t)$,  $\Delta_k=\eps^{(\Lambda_{1,c})}_k-\eps^{(\Lambda_{1,lh})}_k$
is the static L-bandgap energy, $v_k = v^{(\Lambda_{1,c})}_k-v^{(\Lambda_{1,lh})}_k$  is the relative speed between an electron and the light hole
and $V_k (t) = A(t) M_k$ is the matrix element of the transition. The instantaneous eigen-energies are 
\begin{equation}
 \eps_{\pm}(k,t)=\frac{1}{2} (\Delta_k(t) \pm \Delta^{\ast}_k(t)  )
\end{equation}
where $\Delta^{\ast}_k(t) = \sqrt{\Delta^2_k(t) + 4 V^2_k(t)}$ is the renormalized band gap energy. 
For continuous wave laser with $A(t)=F/ \omega_L \sin \omega_L t$, the eigenenergies satisfy the relation  
\begin{equation}
 \eps_{\pm}(k,t) = \eps_{\pm} \left(-k,t \pm m \frac{\pi}{\omega_L} \right) , \quad m=1,3, \ldots 
\end{equation}
with $m$ odd-integer. In this adiabatic approximation, the density matrix of a Bloch state $|k \rangle$
\begin{equation}
\rho_k^{A}=
\left(
\begin{array}{cc}
\sin^2(\theta_k/2) &   \frac{1}{2} \sin \theta_k\\
\frac{1}{2} \sin \theta_k & \cos^2(\theta_k/2)
\end{array}
\right )
\end{equation}
is parametrized by the  mixing angle between the light-hole and the conduction band $\cos \theta_k(t)=\Delta_k(t)/\Delta^{\ast}_k(t)$, 
the transient occupation number   
\begin{equation}
\label{fcad} 
 f_{ck}(t) = \frac{1}{2}  (1-\cos \theta_k(t)),
\end{equation}
describes a continuous population and depopulation of the conduction band. The density of these virtual 
electron-hole pairs $n=\sum_k f_{ck}$  oscillates in time following the periodicity of the incident irradiation,
$n$ vanishes at the extrema of the electric field and attains maximum at the zeros of the electric field. 
For monochromatic laser irradiation with $A(t)=F/\omega_L \sin \omega_L t$, the amplitude of the oscillating  electron density is
\begin{equation}
\label{nmax}
n_{max}= \sum_k \left(1-
\frac{\Delta_k+v_k F/ \omega_L} 
{\sqrt{ (\Delta_k+v_k F/ \omega_L)^2+4M^2_kF^2/\omega^2_L}} \right).
\end{equation}
For a given field strength $F$, the amplitude of the oscillating density $n_{max}$ increases monotonically with the increase of the laser wavelength.
In the regime of strong laser fields $F\rightarrow \infty$ or low-frequency irradiation $\omega_L \rightarrow 0$,
the band gap energy in Eq.(\ref{nmax}) can be neglected, such that the amplitude of the oscillating density becomes independent on the laser parameters
and reaches a saturation limit
\begin{equation}
\label{delta1}
n_{sat} = \sum_k \left(1- \frac{v_k}
{\sqrt{v_k^2+4M^2_k}} \right) , \quad F \rightarrow \infty 
\end{equation}
The population inversion of a state $|k \rangle$, $f_{vk}-f_{ck} = \cos \theta_k$ gives rise to ultrafast intraband  current  
\begin{equation}
 J_{{\rm intra}} = \sum_k (A(t)+v_k \cos \theta_k), 
\end{equation}
following the temporal profile of the laser vector potential, this non-linear AC current does not vanish because Bloch electrons moving with 
quasimomenta $k$ and $-k$ experience different band gap energies in presence of the laser, i.e.
$\Delta_{\pm k}(t)= \Delta_k \pm v_k A(t)$ as a consequence of the sign reversal of the group  velocity $v_{\pm k} =-v_{\mp k}$ 
under the reflection $k \rightarrow -k$. Thus in this adiabatic approximation, the main effect of the laser is to create a transient 
asymmetry in the population inversion of a Bloch state $k$,  which allows the generation of ultrafast and completely reversible 
currents inside the bulk.  Similarly, ultrafast interband current is generated due to the admixture of conduction electron character  
into the light hole band 
\begin{equation}
 J_{{\rm inter}} = \sum_k M_k \sin \theta_k.
\end{equation}  
These transient macroscopic currents produce temporal variation of the electronic energy relative to the ground state energy
\begin{equation}
\label{en}
E(t)-E_0=\sum_k[\Delta_k-\Delta^{\ast}_k(t)+A^2(t)], 
\end{equation}
which includes two distinct contributions: the first term is the difference between the static band gap energy and the 
renormalized bandgap energy, which gives negative contribution to the total energy because of the 
repulsion between the instantaneous valence and conduction band energy levels, the second free-electron Drude term $A^2(t)$ is always positive.  
For moderate laser intensity, the competition of these two effects causes transient fluctuation  of the electronic energy 
relative to the ground state energy. These oscillations exist only during the pulse, but no energy is deposited into the bulk after the end of the pulse.

We further expand the state vector in the adiabatic basis 
\begin{equation}
 |\psi_k(t) \rangle = \alpha_k(t) |-, k, t \rangle + \beta_k(t) | +, k, t \rangle 
\end{equation}
with amplitudes satisfying the equations of motion 
\begin{eqnarray}
 & & i \partial_t \alpha_k(t) = \eps_-(k,t) \alpha_k(t) + F(t) d_k(t) \beta_k(t) \\
 & & i \partial_t \beta_k(t) = \eps_+(k,t) \beta_k(t) + F(t) d_k(t) \alpha_k(t)
\end{eqnarray}
where $d_k(t)=d_k [\Delta_k/\Delta^{\ast}_k(t)]^2$ is the time-dependent transition dipole moment with $d_k=i M_k/\Delta_k$.  
The change in the number density of photoelectrons due to photoionization 
\begin{equation}
\label{den}
\delta n_e(t) = - \sum_k \sin \theta_k(t) \Re [\alpha^{\ast}_k(t) \beta_k(t)] + \sum_k |\beta_k(t)|^2 \cos \theta_k(t)  ,
\end{equation}
includes an interference term due to the field-driven phase coherence between the valence and the conduction band 
(first term),  the second term describes carrier generation due to photionization. 
The transient term vanishes at the zeros of the vector potential and  vanishes long after the end of the pulse $t \rightarrow \infty$. 
The carrier photogeneration term is most effective near the extrema of the electric field when real electron-hole pairs are born,  
the total photoelectron yields after the end of the pulse is $n_e=\sum_k|\beta_k(\infty)|^2$ 
and the absorbed laser energy  is $E(\infty)-E_0  = \sum_k \Delta_k|\beta_k(\infty)|^2$.

To analyze the cycle-resolved electron dynamics and HHG, we treat the non-adiabatic coupling $F(t)d_k(t)$ as weak and neglect the photionization term $\sum_k|d_k|^2$, 
to first order of perturbation theory the interference term is 
\begin{equation}
\label{interf}
 \alpha^{\ast}_k(t) \beta_k(t) =  -i \int_{-\infty}^t dt' e^{-i S_k(t,t')} F(t') d_k(t') 
\end{equation}
where $S_k(t,t')=\int_{t'}^t dt'' \Delta^{\ast}_k(t'')$  
is the quasi-classical action for an electron-hole pair. For monochromatic laser irradiation, 
we expand the phase-factor in Fourier series 
\begin{equation}
e^{i S_k(t,-\infty)} = e^{i \tilde{\Delta}_k t} \sum_n M_n(k)  e^{i n \theta} 
\end{equation}
in terms of the phase angle $\theta=\omega_L t$ with   
\begin{equation}
M_n(k)=\frac{1}{2 \pi} \int_0^{2 \pi} d\theta e^{i S_k(t)}  e^{-i n \theta} e^{-i \tilde{\Delta}_k t}.  
\end{equation}
Expansion of the subintegrand in Eq.(\ref{interf})   
\begin{equation}
 L(k,t)= e^{i S_k(t)} F(t) d_k(t)= e^{i \tilde{\Delta}_k t} \sum_n L_n(k) e^{i n \theta}  
\end{equation}
with 
\begin{equation}
L_n(k)=\frac{1}{2 \pi} \int_0^{2 \pi} d \theta e^{i S_k(t)} F(t) d_k(t) e^{-i n \theta} e^{-i \tilde{\Delta}_k t}, 
\end{equation}
gives the Fourier series of the interference term  
\begin{equation}
 \alpha^{\ast}_k(t) \beta_k(t) = \sum_l e^{i l \theta} f_{kl}  
\end{equation}
where 
\begin{equation}
\label{fkn} 
f_{kl}=\sum_n M^{\ast}_{n-l}(k) \frac{1}{\tilde{\Delta}_k+n \omega_L+i \eta } L_n(k)  
\end{equation}
here $\eta>0$ is a positive infinitesimal. For each fixed harmonic order $l$, the Fourier coefficients in Eq.(\ref{fkn}) describe interference between 
multiphoton transitions (both energy-conserving and virtual). The terms with $n<0$ correspond to multiple absorption of laser photons,
while terms with $n>0$ correspond to stimulated emission of photons.  

For moderate laser intensity considered in the present work, we approximate the energy-momentum dispersion curve with 
\begin{equation}
\Delta^{\ast}_k(t) \approx \Delta_k + v_k A(t) + \frac{A^2(t)}{2 m^{\ast}_k} 
\end{equation}
here $m^{\ast}_k$ is the effective mass of an electron-hole pair $1/m^{\ast}_k= 4 M_k^2/\Delta_k$ (cf. also Ref.\cite{Ashcroft}). Using the Jacobi-Anger expansion 
\begin{equation}
 e^{ i x \cos \theta} e^{-i y \sin 2 \theta}=\sum_n i^n e^{i n \theta} J_n(x,y) , 
\end{equation}
in terms of the generalized Bessel function \cite{Reiss} 
\begin{equation}
 J_n(x,y)=\sum_l J_{n-2l}(x) J_l(y) 
\end{equation}
where $J_n(x)$ are ordinary Bessel functions, we obtain 
\begin{equation}
 M_n(k) = i^n J_n(x_k,y_k)  
\end{equation}
with intensity parameters
\begin{equation}
 x_k = \xi v_k, \quad y_k =\frac{U_k}{ 2 \omega_L}
\end{equation}
where $\xi= F/\omega_L^2$ is the excursion length for a free electron and $U_k=F^2/4 m^{\ast}_k \omega_L^2$ is the pondermotive energy of an electron-hole pair
in the effective mass approximation.  Making an approximation 
\begin{equation}
 L_n(k) \approx i^{n+1} [J_{n+1}(x_k,y_k)-J_{n-1}(x_k,y_k)] \frac{F \tilde{d}_k}{2}
\end{equation}
in terms of cycle-averaged transition dipole moment $\tilde{d}_k$, we obtain 
\begin{equation}
 f_{kl} = \frac{F \tilde{d}_k}{2} 
 i^{l+1}  \sum_n J_{n-l}(x_k,y_k) \frac{1}{\tilde{\Delta}_k+n \omega_L + i \eta} [J_{n+1}(x_k,y_k)-J_{n-1}(x_k,y_k)]
\end{equation}
with effective bandgap energy $\tilde{\Delta}_k= \Delta_k+U_k$.

\begin{figure}
\begin{center}
\includegraphics[width=.85\textwidth]
{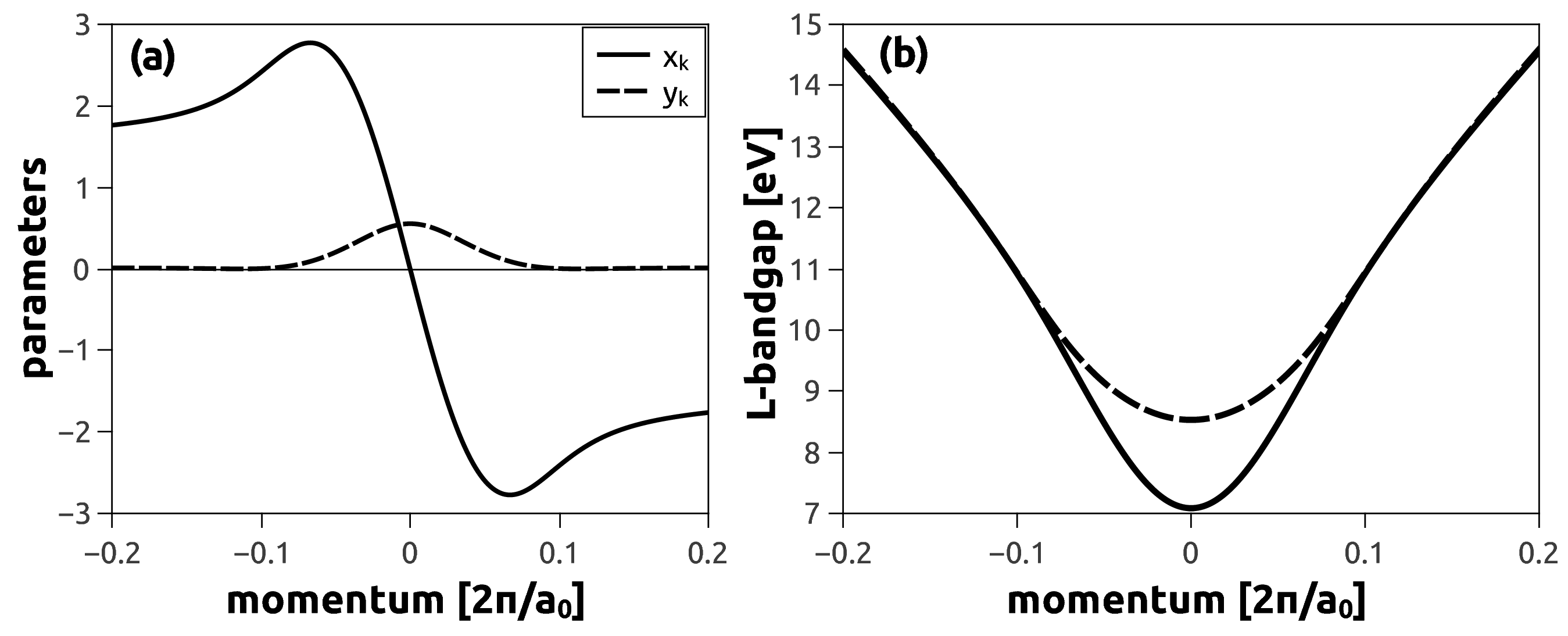} \caption{(a) Momentum dependence of the intensity parameters $x_k=\xi v_k$ and $y_k=U_k/2 \omega_L$, where $\xi=F/\omega_L^2$ is the excursion length 
and $U_k$ is the pondermotive energy of an electron hole pair. (b) Solid line - undistorted bandgap energy $\Delta_k$, 
and dashed line - cycle-averaged renormalized bandgap energy $\tilde{\Delta}_k$.} \label{fig03}
\end{center}
\end{figure}

Figure \ref{fig03}(a) shows the crystal-momentum dependence of the intensity parameters $(x_k,y_k)$ for the peak laser field strength 
$F=$ 0.4 V/{\AA} and photon energy $\hbar \omega_L=1.55$ eV. For these laser parameters, the excursion length in the oscillatory motion of a free electron $\xi=2.4$ 
is smaller than the lattice constant,  the parameter $x_k=\xi v_k$ gives the scaled velocity distribution of electron-hole pairs.
Because of the inhomogeneous distirbution of the pondermotive energy in the Brillouin zone,  
electron-hole pairs with crystal momentum  near the Brillouin zone center experience strong pondermotive energy shifts with $U_k \ge \omega_L$, 
while states with $|k| > 0.1$ are not affected. As a consequence the effective bandgap energy $\tilde{\Delta}_k$, shown in Fig. \ref{fig03}(b),
is strongly renormalized at the Brillouin zone center: the minimal 
energy needed to create a real electron-hole pair increases to 8.4 eV.  Thus for these specific parameters of the laser field, 
the 5-photon threshold for photoionization is closed, and above threshold ionization with absorption of 6 laser photons is energetically allowed.

\begin{figure}
\begin{center}
\includegraphics[width=.5\textwidth]
{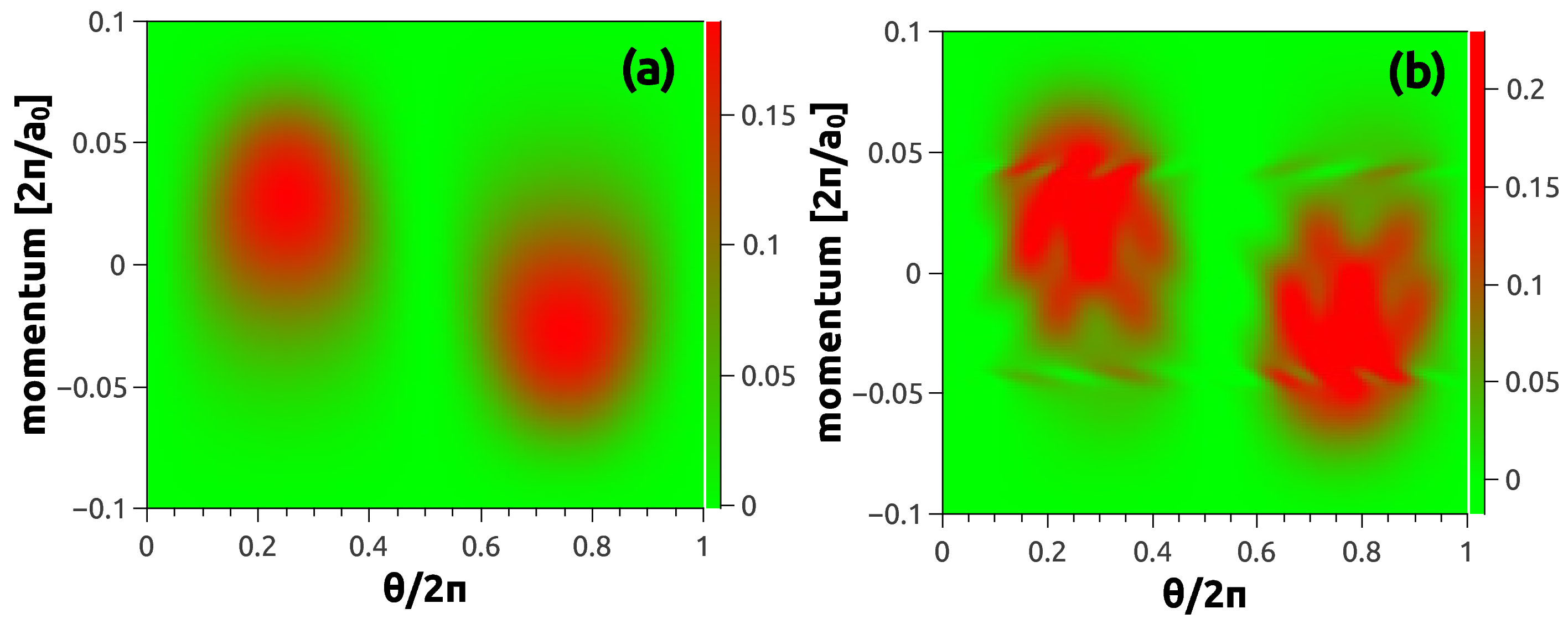} \caption{Transient electron distribution function in bulk diamond along the $\Lambda$ line in the Brillouin zone for
one period of oscillation of the electric field of a continuous wave laser having a photon energy $\hbar \omega_L=1.55$ eV and field strength $F=0.4$ V/{\AA},
the phase parameter is $\theta=\omega_L t$. Fig. (a) - density distirbution in the adiabatic approximation and Fig.(b) - the density distribution including 
the non-adiabatic effects (cf. also text). 
} \label{fig04}
\end{center}
\end{figure}

Figure \ref{fig04}(a) shows the transient distribution function $f_e(k,t)$ of electron-hole pairs in the adiabatic approximation, cf. Eq.(\ref{fcad}).
The adiabatic density distribution follows the temporal profile of the laser vector potential, the distribution is concentrated in the range  with $|k| < 0.1$
and exhibits charactersitic direction dependence following the symmetry trend $f_e(k,t)=f_e(-k,t \pm m \pi/\omega_L)$ with $m$ odd-integer. 
During the first half-cycle, electron-hole pairs with velocities $v_k < 0$ in the range $k>0$ 
experience contraction of the transient bandgap energy $\Delta_k(t)=\Delta_k-|v_k| A(t)$, which enables hybridization of the valence 
with the conduction band and thus enhances the  number of conduction electrons in that range. During the second half-cycle, electron-hole pairs   
in the range $k <0$ moving with reversed speeds $v_k > 0$ experience contraction of the band gap energy and thus these valence electrons transiently emerge  into the conduction band. 
This movement of charge carriers results in transient asymmetry of the population inversion, i.e. at each instant of time $t$  there are either more or less electrons 
in a state $|k\rangle$ than in the state $|-k\rangle$, which in turn produces completely reversible non-linear interband current.

\begin{figure}
\begin{center}
\includegraphics[width=.5\textwidth]
{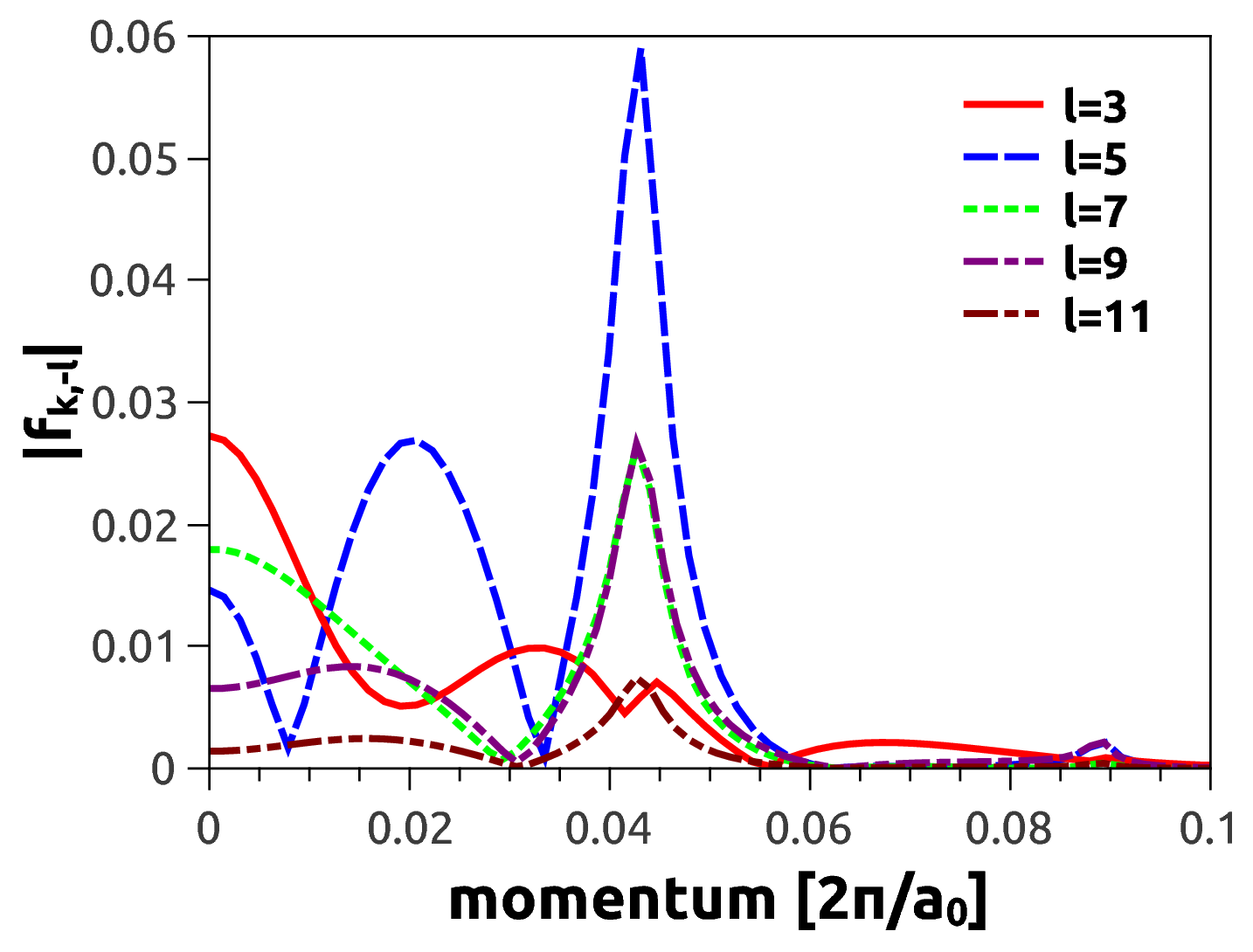} \caption{Strength of laser-induced odd-order harmonics in diamond bulk along the $\Lambda$ line in the Brillouin zone. 
The crystal is subjected to continuous wave
laser with photon energy $\hbar \omega_L=1.55$ eV, the electric field strength is $F=0.4$ V/{\AA}} \label{fig05}
\end{center}
\end{figure}

Figure \ref{fig04}(b) shows the density distribution including the non-adiabatic response of electrons to the laser electric field. 
The adiabatic time evolution is distorted and rapid sub-cycle structure emerges associated with generation of high order harmonics. 
The highly energetic modulation of the photoelectron density is localized near the threshold wave-numbers $k_6=\pm 0.04$ 
associated with resonant 6-photon transition through the renormalized bandgap with $\tilde{\Delta}_k=6 \omega_L$.  
The  strength of density fluctuations associated with HHG $|f_{k,-l}|$ with $l$  odd positive integer is exhibited in Fig.\ref{fig05}. 
The interference term (cf. Eq.(\ref{interf}) produces intense 5th harmonic near the 6-photon threshold, 
the 7th and 9th harmonics are of equal amplitude near the threshold wave-number and plateau region is exhibited. 
The intensity of the fluctuating electron density falls off rapidly when $l>9$, thus for the specific intensity parameters considered in this work $l=9$ specifies a cutoff energy for HHG.

\subsection{Multiband model}

In the multiband model based on Eqs.(\ref{eom}), the Brillouin zone (BZ) was sampled by a Monte Carlo method using 3000 randomly
generated $\ks$-points, $N=$20 bands were included in the expansion of the wave-packet over Bloch orbitals, with 4 valence and 16 conduction bands.
The wave-functions of initially occupied valence band states were propagated forward in time by applying the Cayley transformation
for small equidistant time steps $\delta=0.03$ a.u.
\begin{equation}
 |\psi_{v \ks}(t+\delta) \rangle = [1+i  H(\ks,t)\delta/2]^{-1} [1-i  H(\ks,t)\delta/2]  |\psi_{v \ks}(t)\rangle 
\end{equation}
with $H(\ks,t)=\exp(i \ks \cdot \rs) H(t) \exp(-i \ks \cdot \rs)$.

\begin{figure}
\begin{center}
\includegraphics[width=.5\textwidth]
{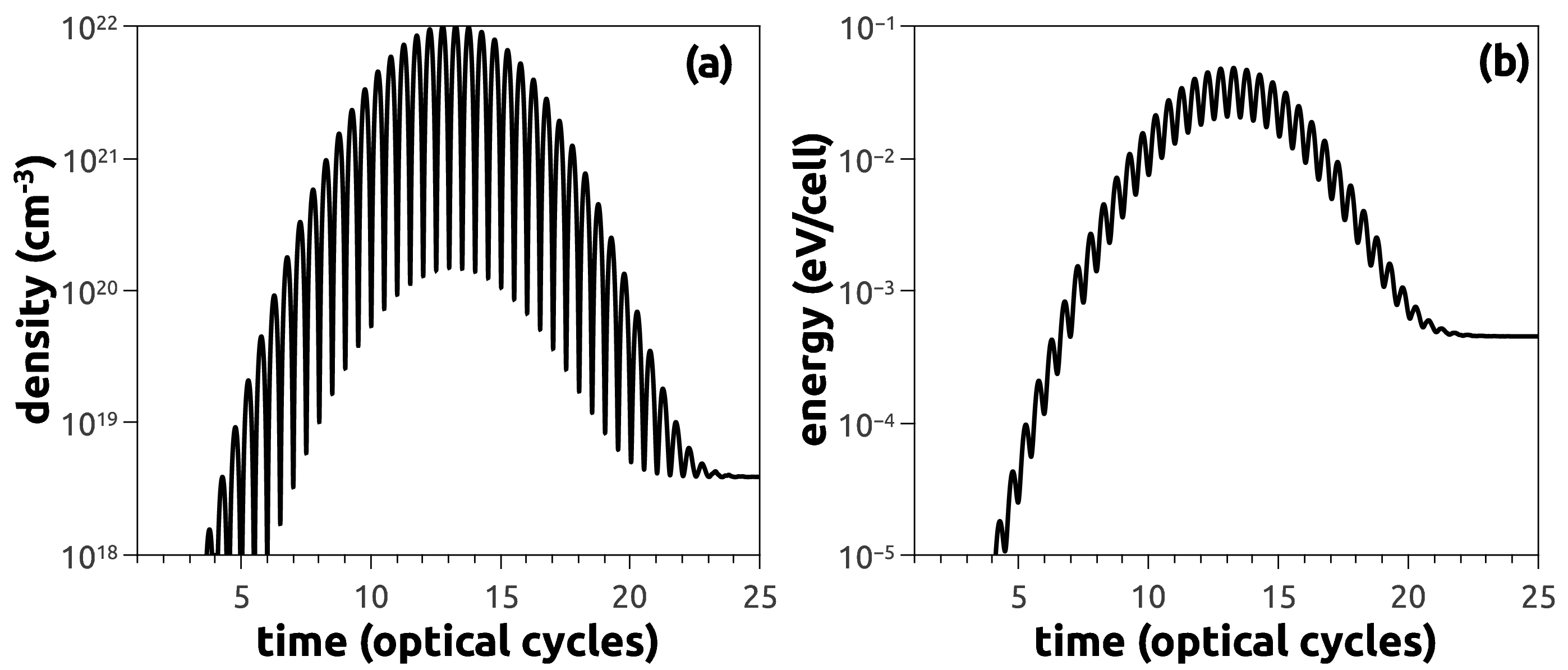} \caption{(a) Time evolution of the electron density in diamond bulk
irradiated by 15fs laser pulse with peak field strength  0.4 V/{\AA},
linearly polarized along the $\Lambda$ line in the Brillouin zone and (b) Time developement of the electronic excitation energy per unit cell.
In Fig.(a-b) the time interval is measured in optical cycles of the laser field, one optical cycle has a duration of 2.75 fs} \label{fig1}
\end{center}
\end{figure}

In Fig. \ref{fig1}(a-b), we plot the time evolution of the electron density and the excitation energy per unit cell for peak laser field strength $F=0.4$ V/{\AA}.
The total number of electron-hole pairs increases gradually on the rising edge of the pulse. Charge carriers are born at the extrema of the electric field,
their number density increases rapidly during
the first quarter of each cycle and attains maximum at the time when the electric field reverses direction, subsequently electrons
recombine with their corresponding valence band holes during each second quarter of the cycle, resulting in transient charge density oscillations,
following the adiabatic time evolution. After the peak intensity, large fraction of virtual carriers recombine, such that the
cycle-averaged electron yield gradually decreases and stabiles in the wake of the pulse. The electronic excitation energy is shown in Fig.\ref{fig1}(b):
the energy gain increaes  gradually on the rising edge of the pulse due to multiphoton absorption, the temporal fluctuation of the absorbed energy 
is negligible, the cycle-averaged energy deposited onto electrons reaches 0.1 eV per unit cell at the pulse peak,
well below the cohesive energy of diamond (about 7 eV/atom). Because large fraction of the virtual electron-hole pairs disappears after the pulse peak, the  
final energy gain is substantially reduced to $10^{-3}$ eV/unit cell.

\begin{figure}
\begin{center}
\includegraphics[width=.5\textwidth]
{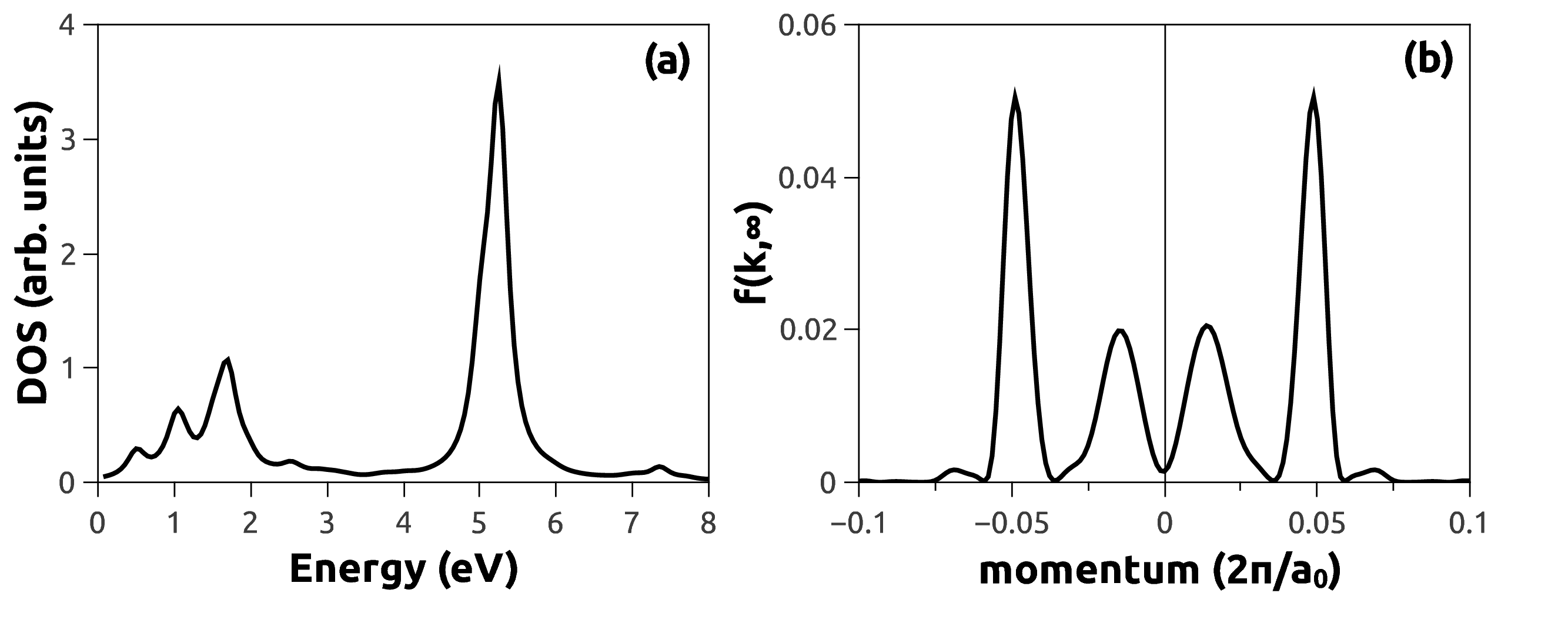} \caption{(a) Density of conduction states after the irradiation of
bulk diamond with 15fs laser pulse, having a peak intensity 0.4 V/{\AA},
linearly polarized along the $\Lambda$ line in the Brillouin zone; energy is measured
relative to the conduction band minimum. (b) Distribution function of electrons along the direction of laser polarization;
the quasimomentum is measured in units $2 \pi/a_0$, where
$a_0=$3.57 {\AA}  is the bulk lattice constant.} \label{fig2}
\end{center}
\end{figure}

In Fig.\ref{fig2}(a-b), we plot the density of conduction states and the electron distribution function after the end of the pulse.
Photoelectrons are excited across the direct bandgap at the $\Gamma$ point and occupy the $\Gamma_{2'}$ band with energies 2 eV above the conduction band minimum.
Noticeably however substantial fraction of carriers are promoted into  the $\Gamma_{15}$ band with energies 5 eV above the conduction band minimum (cf.  Fig.\ref{fig2}(a)) .
The asymptotic momentum distribution shown in Fig.\ref{fig2}(b) is almost symmetric under inversion with $\ks \rightarrow -\ks$,
electrons occupy states with quasi-momentum in a narrow range near the BZ center. Two groups of occupied conduction states can be distinguished:
states in the $\Gamma_{15}$ band produce the inner peaks of the distribution, while the states in the outer two peaks with $k=\pm 0.05$
are due to excitation into the $\Gamma_{15}$ band. While occupation of the $\Gamma_{2'}$ band is due to interband transition $\Gamma_{25'} \rightarrow \Gamma_{2'}$,
the population of the $\Gamma_{15}$ band is primary due to indirect interaction with $\Gamma_{2'}$ band. That is because the electric field breaks the symmetry of
the diamond lattice under spatial inversion and as result the two conduction bands are mixed with the valence band $\Gamma_{25'}$, which enables transfer of population
from the $\Gamma_{2'}$ to $\Gamma_{15}$ band during the irradiation. The comparison of this result with the prediction of our simplified two band model
including the coupling between the light hole and the lowest conduction band, shows that the effect of interaction between multiple conduction bands
can not be neglected for the photoionization of diamond bulk.

\begin{figure}
\begin{center}
\includegraphics[width=.5\textwidth]
{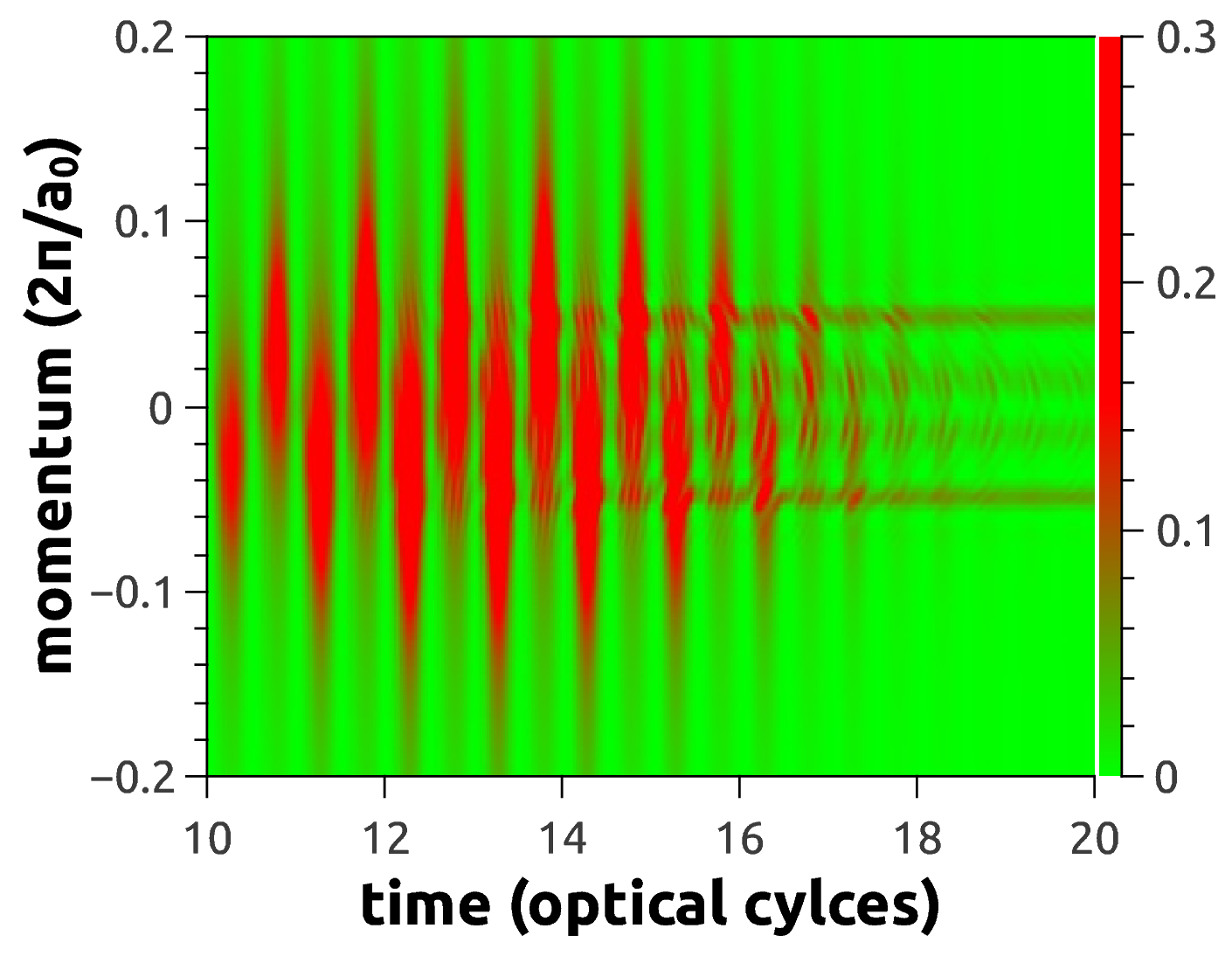} \caption{Time evolution of the electron distribution function in bulk diamond along the $\Lambda$ line in the Brillouin zone.
The time interval is measured in optical cycles of the laser field, one optical cycle has a duration of 2.75 fs.
The crystal momentum is measured in units $2 \pi/a_0$, where $a_0=$3.57 {\AA}  is the bulk lattice constant. } \label{fig3}
\end{center}
\end{figure}

The time evolution of the phase-sensitive electron distribution function $f_e(\ks,t)$ is shown in Fig. \ref{fig3}(a-b).
The laser excites photoelectrons close to the BZ center because of the localized character of the interactions between the light hole and  conduction bands
(cf. Fig.(\ref{fig02}). During each two adjacent half-cycles of the driving field, the distribution exhibits the symmetry trend
\begin{equation}
 f_e(\ks,t) \approx f_e \left(-\ks,t \pm \frac{\pi}{\omega_L} \right), \label{dyn_symf}
\end{equation}
because for monochromatic laser irradiation $F(t)=F \cos \omega_L t$ the Hamiltonian remains invariant under the combined transformation
of spatial inversion $\rs \rightarrow -\rs$ supplemented by  time translation by half of the laser period $t \rightarrow t \pm \pi/\omega_L$. For pulsed
laser irradiation this discrete symmetry is  broken by the finite pulse duration $\tau$. Nevertheless remnants of that symmetry  are prominent, so that electrons
move rapidly back and forth about the BZ center, following the adiabatic time evolution.  Electron-hole pairs are born at the extrema
of the laser electric field and their occupation number increases dramatically during the first quarter of each cycle. These virtual pairs
pop-out of existence during the second quarter of the cycle, at the time when the field reverses direction. During the third quarter of the cycle,
electrons re-emerge in the conduction band following the  trend of Eq. (\ref{dyn_symf}).
Sub-cycle structure and fine details in the transient density distribution are exhibited when the intensity of the laser increases, 
in excellent agreement with the prediction of the two-band model. The asymptotic distribution ($t \rightarrow \infty$) 
of real electron-hole pairs is structured into four narrow peaks symmetrically displaced about the BZ center.

\begin{figure}
\begin{center}
\includegraphics[width=.5\textwidth]
{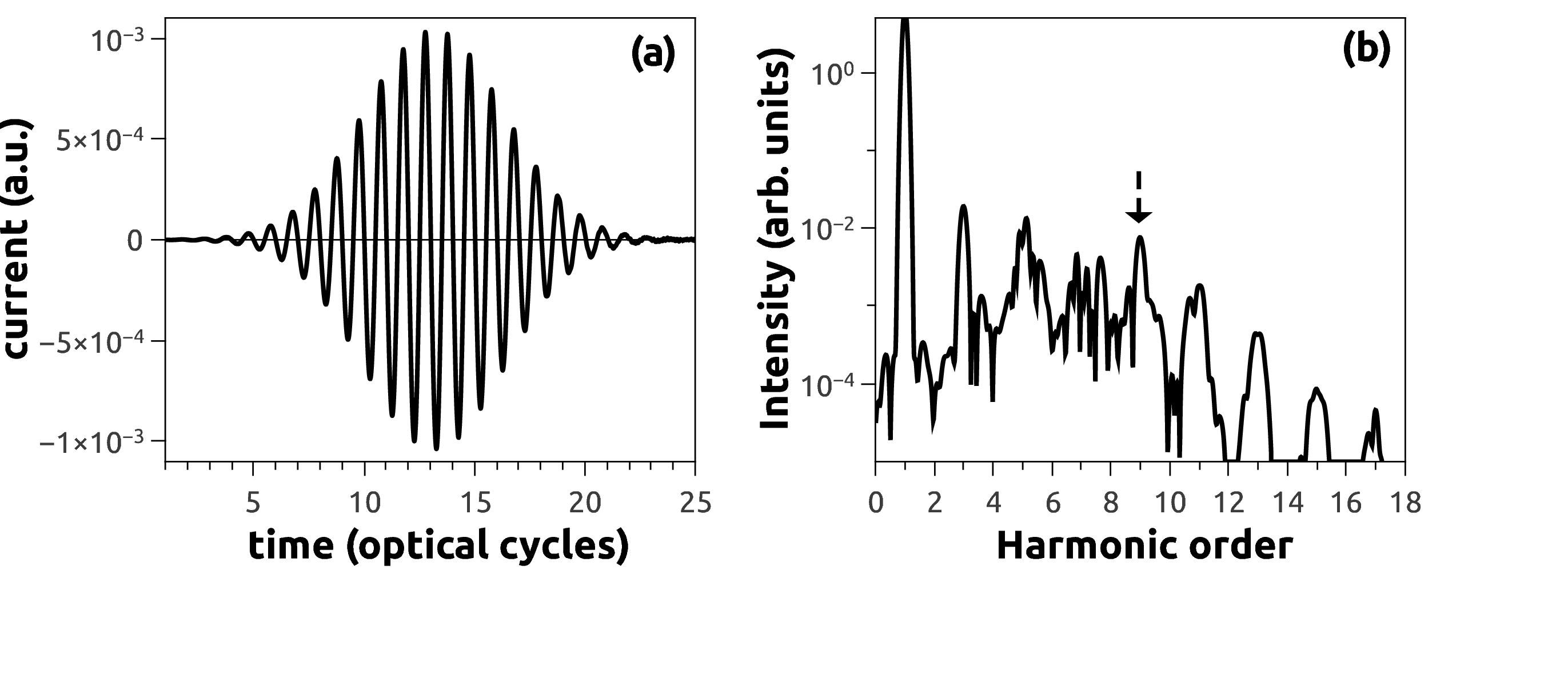} \caption{(a) Time evolution of the optically-induced electronic current in bulk diamond irradiated by
15fs laser pulse of wavelength 800 nm, having peak field strength 0.4 V/{\AA} and
linearly polarized along the $\Lambda$ line in the Brillouin zone and (b) the corresponding spectrum of high-order harmonics.} \label{fig4}
\end{center}
\end{figure}

The time-development of the HHG current is also shown in Fig.\ref{fig4}(a), together with the corresponding HHG spectrum in Fig.\ref{fig4}(b).
The laser induces reversible ultrafast currents  following the temporal profile of the laser vector potential.
The HHG spectrum is characterized by odd harmonics structure with plateau region and sharp cutoff at the position of the 9th harmonic.
In fact, the position of the cutoff is in good quantitative  agreement with the prediction of the two band model, because the strength of the transient density
fluctuations is substantially reduced when $l > 9$. The suppression of even order harmonics is a consequence of the approximate symmetry trend 
$J(\ks,t) \approx -J \left(-\ks,t \pm \pi/\omega_L \right)$, which implies that all even order harmonics interfere destructively.

\begin{figure}
\begin{center}
\includegraphics[width=.5\textwidth]
{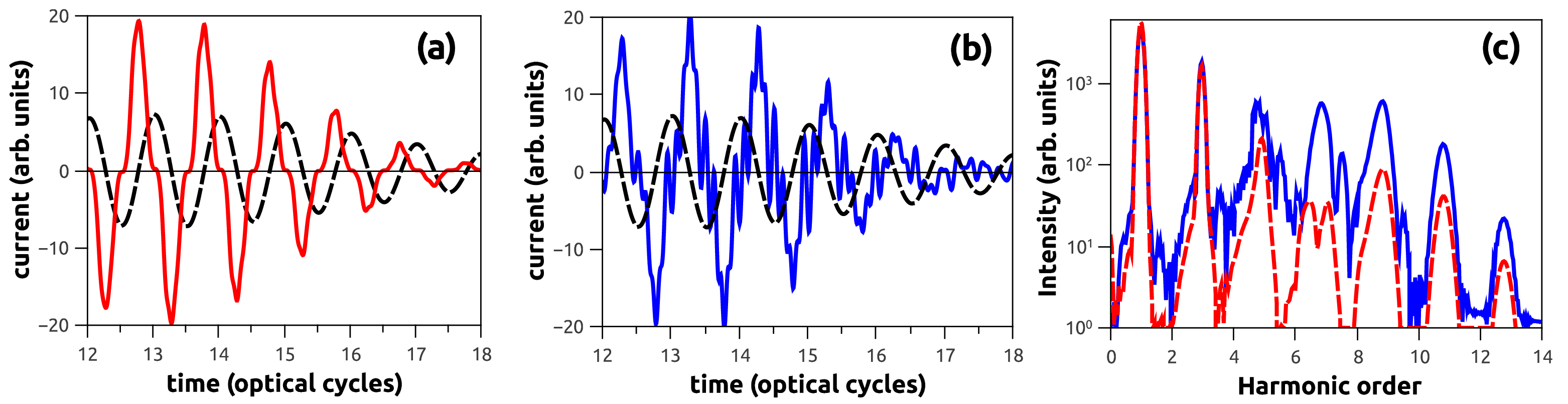} \caption{(a) Time evolution of the optically-induced intraband current in the diamond bulk along the $\Lambda$ line in the Brillouin zone and Fig. (b)
gives the transient interband current along the same line. The dashed line in Fig.(a-b) indicates the temporal profile of the laser electric field
and Fig.(c) inlcudes the spectrum of the corresponding intraband (dashed line) and interband harmonics (solid line). } \label{fig5}
\end{center}
\end{figure}

The total current can be conventionally split into intra- and interband currents \cite{gauge}. Since HHG is a non-linear process,
the spectrum of high-order harmonics is not affected by the transformation $J(t) \rightarrow J(t)+ const \times J_D(t)$,
where $J_D(t)$ is the Drude  current due to collective plasma oscillation of the electron gas.
This splitting allows us to expose the relevant non-linear effects associated with intra- and interband motion of charge carriers.
In Fig.\ref{fig5}(a) we show the intraband current along the $\Lambda$ line, by excluding the Drude current.
Fig.\ref{fig5}(b) shows the interband current shifted by the Drude current. The currents are almost completely reversible and follow the periodicity of
the laser vector potential. The intraband current vanishes at the extrema of the electric field and exhibits stationary inflection points; 
that is because near the extrema of the electric field, charge carriers decelerate in their respective bands and reverse their speeds, during this transient phase their kinetic energy
can be converted into a photon, quite similarly to the bremsstrahlung process. High-order harmonics are emitted in the diamond bulk
during each half-cycle with alternating phase (i.e. only odd order harmonics are emitted).  Similarly the interband current follows
the period of the laser vector potential, but flows in direction opposite to the intraband current. The rapid sub-cycle structure exhibited in the interband current
is a direct consequence of the build-up of coherent superposition of populations in the valence and conduction bands.
During each half-cycle, the rapidly oscillating dipole emits radiation in the form of high-order harmonics. Fig.\ref{fig5}(c) shows
that for the specific 800 nm wavelength, interband harmonics are much more intense in the plateau and cutoff regions, while the intraband current has
prominent contribution in the generation of sub-bandgap harmonics.

\section{Conclusion}

In summary, we have investigated the quasi-adiabatic sub-cycle dynamics of electron-hole pairs and the high harmonic generation
in bulk diamond subjected to intense ultrashort laser pulse. For near-infrared laser having wavelength 800 nm, pulse duration 15fs and peak field strength 0.4 V/{\AA}
we have compared the relative importance of two main mechanisms of high harmonic generation: intraband emission at each half-cycle of the laser field,
which arises due to transient asymmetry in the population inversion of adibatically evolved Bloch states and
interband emission arising due to build-up of coherent superposition of populations in the valence and  conduction bands. The HHG spectrum of the interband current dominates the plateau region of HHG,
while the intraband current has primary contribution in the sub-bandgap region. The modification of the substrate band structure
occurs on a sub-femtosecond time scale, transforming diamond into a transiently metallic state supporting non-linear ultrafast currents that have 
definite phase relation with respect to driving laser field and generate high order harmonics during each half cycle of the pulse.

\section*{Acknowledgements}
This work is supported by the Bulgarian National Science Fund under Contracts No.DFNI-E02/6, No. DNTS/FRANCE-01/9,(T.A.) and by the Bulgarian National Science Fund under Contract No. 08-17 (B.O.).

\end{document}